\documentclass[conference]{IEEEtran}
\IEEEoverridecommandlockouts
\usepackage{cite}
\usepackage{amsmath,amssymb,amsfonts}
\usepackage{algorithmic}
\usepackage{graphicx}
\usepackage{textcomp}
\usepackage{xcolor}
\usepackage{physics}
\usepackage{cuted}

\usepackage{siunitx}
\usepackage{comment}
\usepackage{orcidlink}

\newcommand{\figref}[1]{Fig.\,\ref{#1}}
\newcommand{\Figref}[1]{Figure\,\ref{#1}}
\newcommand{\secref}[1]{Sec.\,\ref{#1}}
\newcommand{\tabref}[1]{Table\,\ref{#1}}

\def\BibTeX{{\rm B\kern-.05em{\sc i\kern-.025em b}\kern-.08em
    T\kern-.1667em\lower.7ex\hbox{E}\kern-.125emX}}

\begin{document}

\title{Model Predictive Online Trajectory Planning for Adaptive Battery Discharging in Fuel Cell Vehicle 
}

\author{
\IEEEauthorblockN{Katsuya Shigematsu, Hikaru Hoshino, Eiko Furutani}
\IEEEauthorblockA{\textit{Department of Electrical Materials and Engineering} \\
\textit{University of Hyogo}\\
2167 Shosya, Himeji, Hyogo 671-2280, Japan \\
er23o008@guh.u-hyogo.ac.jp, \{hoshino, furutani\}@eng-u-hyogo.ac.jp}
}


\maketitle

\begin{abstract}
This paper presents an online trajectory planning approach for optimal coordination of Fuel Cell (FC) and battery in plug-in Hybrid Electric Vehicle (HEV). 
One of the main challenges in energy management of plug-in HEV is generating State-of-Charge (SOC) reference curves by optimally depleting battery under high uncertainties in driving scenarios.  
Recent studies have begun to explore the potential of utilizing partial trip information for optimal SOC trajectory planning, but dynamic responses of the FC system are not taken into account. 
On the other hand, research focusing on dynamic operation of FC systems often focuses on air flow management, and battery has been treated only partially. 
Our aim is to fill this gap by designing an online trajectory planner for dynamic coordination of FC and battery systems that works with a high-level SOC planner in a hierarchical manner. 
We propose an iterative LQR based online trajectory planning method where the amount of electricity dischargeable at each driving segment can be explicitly and adaptively specified by the high-level planner. 
Numerical results are provided as a proof of concept example to show the effectiveness of the proposed approach. 
\end{abstract}


\section{Introduction}

Hydrogen Fuel Cell Vehicles (FCVs) have emerged as a promising solution in the automotive industry for reducing greenhouse gas emissions. 
Especially, when considering heavy-duty applications, FCVs offer distinct advantages over battery electric vehicles (BEVs) due to large mileage and cost-effectiveness when operating at full payload capacity~\cite{Sagaria2021}.
FCVs usually have a hybrid configuration with a Fuel Cell (FC) system and a battery storage system, which overcomes the slow dynamic response of the FC system~\cite{Amin2014}. 
With plug-in Hybrid Electric Vehicles (HEVs), grid electricity charged to the battery can be consumed during a trip. 
The so-called charge-depleting/charge-sustaining strategies~\cite{Zhang2011} involve discharging the battery initially to a certain level of State-of-Charge (SOC) and then maintaining the SOC around that level in the remaining operation. 
However, the myopic battery depletion is not optimal, and devising an appropriate energy management strategy is crucial for optimal coordination of FC and battery storage systems to achieve high fuel economy. 

Various methods have been proposed for determining SOC reference curves. 
Linear or affine decrease of SOC with distance traveled is used in, e.g.,~\cite{Lancandia2013,Gao2021}, and SOC reference curves are built using historical driving data in, e.g.,~\cite{Zhou2020,Li2022}. 
These methods do not require online trip information.
In turn, they may not achieve optimal battery usage due to the lack of trip details. 
With advancements in intelligent transportation systems and navigation technology, it has become feasible to obtain partial trip information such as route segment lengths and average speeds. 
Thus, recent studies have begun to explore the potential of utilizing these trip information to predict future driving cycles and to generate SOC reference curves~\cite{Wei2023,Piras2024}. 
However, how to effectively and efficiently exploit  partial trip information for online energy management of plug-in HEV is an open problem for further investigation. 

In this paper, we discuss how to construct a hierarchical energy management structure that is suitable for adaptive SOC trajectory planning for plug-in HEVs under high uncertainties in driving scenarios. 
Above mentioned approaches \cite{Wei2023,Piras2024} using partial trip information are promising, but these studies utilize a variant of Equivalence Consumption Minimization Strategy (ECMS), which does not consider dynamic responses of FC systems and only ensures a near-optimal local solution~\cite{Piras2024}. 
On the other hand, for the purpose of dynamic optimization of FC systems, air flow controllers are designed using various methods such as adaptive control~\cite{Han2017}, sliding-mode control~\cite{Liu2019}, and Model Predictive Control (MPC)~\cite{Arce2010,Ouyang2017,Neisen2020}.  
However, these methods do not directly optimize battery usage. 
Thus, to construct an efficient energy management strategy that harnesses dynamic coordination of FC and battery systems in FCVs, an appropriate trajectory planner needs to be designed to work in the middle of an upper-level SOC planner and lower-level feedback controllers (see \secref{sec:model} for the proposed hierarchical strategy).

The main contribution of this paper is to propose an online trajectory planning method where the amount of electricity dischargeable in each driving segment can be explicitly and adaptively specified to facilitate generating SOC references in the upper-level SOC planner. 
The proposed method has the following advantages:
\begin{itemize}
    \item We utilize a state-of-the-art online trajectory planning method called iterative Linear Quadratic Regulator (iLQR)~\cite{Li2004,Tassa2012,Howell2019}, which has been developed for robot motion planning and solves an optimal control problem at each time step in an MPC (receding horizon) framework. While many MPC-based energy management strategies are proposed in, e.g.,\cite{Quan2021,Zhou2022,Anbarasu2022,Jia2023}, most of them use highly simplified models to balance with computational burden. This paper uses a nonlinear dynamical model that captures air flow transients in FC systems leveraging computationally efficient algorithm of iLQR. 
    \item We carefully formulate an optimal control problem such that the amount of electricity dischargeable in each driving segment can be specified to facilitate coordination between the upper-level planner. While several studies have addressed dynamic coordination of FC and battery/ultracapacitor systems~\cite{Vahidi2006,Pereira2021}, they focus on charge-sustaining behavior in non plug-in HEVs and coordination with the upper-planner is not intended. 
\end{itemize}

The paper is organized as follows. 
In \secref{sec:model}, a mathematical model of the hybrid FC-battery system is introduced. 
The proposed hierarchical energy management architecture and the online trajectory planning approach are described in \secref{sec:method}. 
Numerical results are provided in \secref{sec:simulation}. 
Conclusions and future works are summarized in \secref{sec:conclusion}.

\section{Model of Fuel Cell and Battery Systems} \label{sec:model}

\begin{figure}
    \centering
    \includegraphics[width=0.96\linewidth]{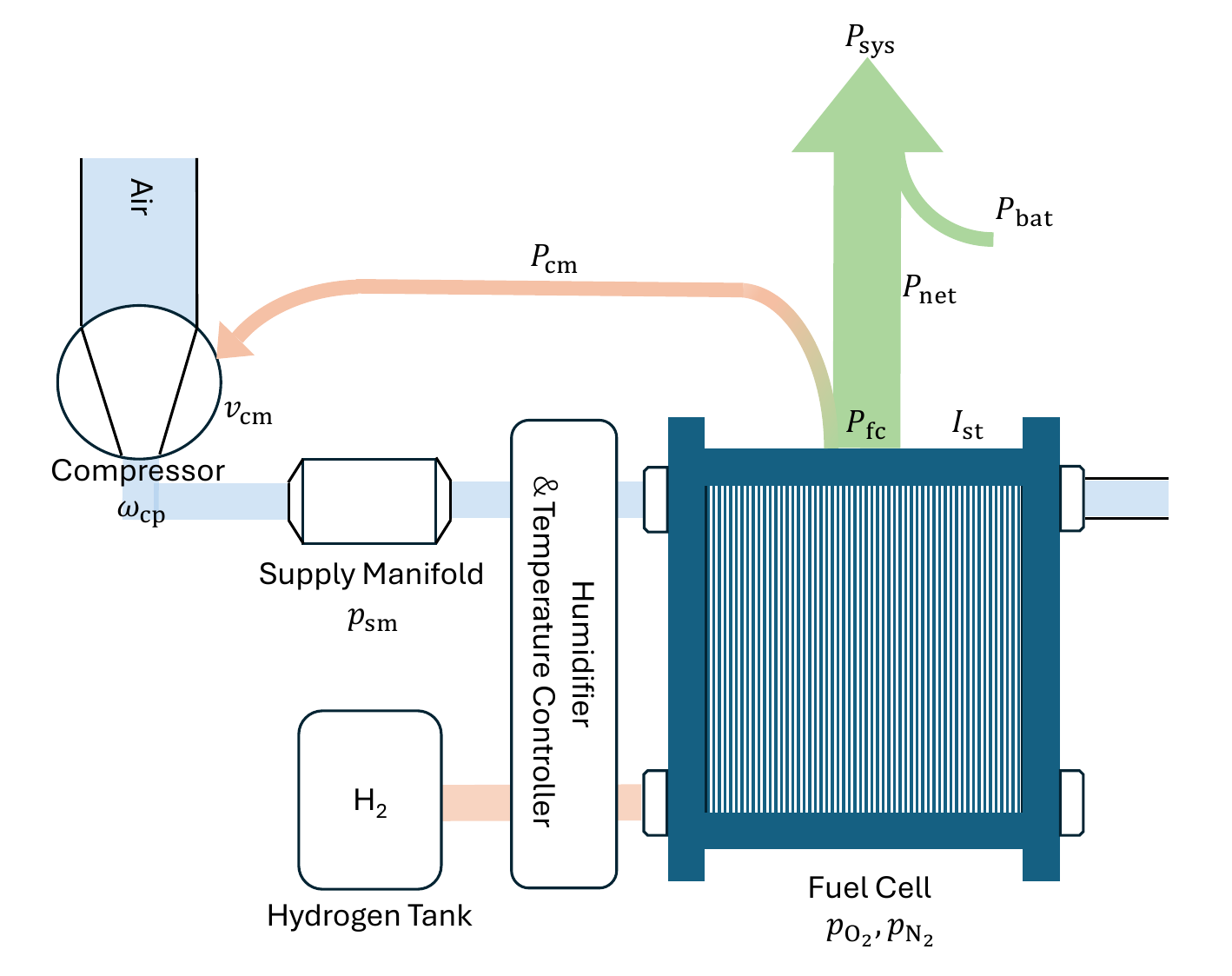}
    \caption{Schematic diagram of the fuel cell system}
    \label{fig:fc_system}
\end{figure}

This section introduces the plant model used for the design of the proposed online trajectory planning method. 
\Figref{fig:fc_system} schematically shows the structure of the FC system, which consists of a Proton Exchange Membrane (PEM) FC stack, 
the air supply subsystem including a compressor and a supply manifold, the hydrogen supply subsystem, and the humidifier and the cooling controller~\cite{Pukrushpan2004}. 
Among them, the air supply system, which provides oxygen to the PEMFC stack, is crucial for the purpose of energy management, since air compressors can consume up to $\SI{30}{\%}$ of the fuel cell power during rapid increase in the air flow~\cite{Vahidi2006}. 
More importantly, there is a challenge that oxygen could be depleted when the stack current increases rapidly.
This oxygen starvation results in a rapid drop of the stack voltage and even reduces the lifetime of the fuel cell membrane. 
However, the air supply rate is limited by the supply manifold dynamics, and centrifugal compressors are susceptible to surge and choke which limits the efficiency and performance of the compressor~\cite{Vahidi2006}. 
One way to avoid oxygen starvation and to match an arbitrary level of current demand is to split the current demand with a battery, which can be connected with the FC system through a DC/DC converter.

In this study, we use a control-oriented model of PEMFC systems proposed in~\cite{Suh2006}.
This model describes essential dynamics 
of the air flow into the cathode of the PEMFC. 
It is assumed that a fast proportional–integral (PI) controller ideally regulates the hydrogen flow to the anode to match the oxygen flow, which means that the hydrogen is supplied from a compressed tank timely and a quasi-steady state is achieved. 
It is also assumed that humidity and temperature are regulated to their desired levels, and the model does not consider the effect of temperature or humidity fluctuations. 
This assumption should not limit the validity of our approach since the temperature and humidity dynamics are considerably slower than the FC power dynamics, and the variation of the temperature and humidity can be captured as a slow variation in model parameters. 

For the design of the proposed online trajectory planner, the dynamics of the FC system combined with a battery storage system are represented by a nonlinear state-space model: 
\begin{equation} \label{eq:state_equation}
\dot{x} = f(x,u)
\end{equation}
where the state vector $x$ is defined as  
\begin{align}
  x &= [x_1, x_2, x_3, x_4, x_5, x_6, x_7, x_8]^\top \notag \\  
         & 
         := [ 
            \underbrace{ p_\mathrm{O_2}, p_\mathrm{N_2} ,\omega_\mathrm{cp},p_\mathrm{sm}}_{x_\mathrm{fc}},  
            \underbrace{ v_\mathrm{soc}, v_\mathrm{s}, v_\mathrm{f} }_{x_\mathrm{bat} },  
            q_\mathrm{dis}
         ]^\top, 
\end{align}
where $x_\mathrm{fc}$ and $x_\mathrm{bat}$ stand for the state vectors for the FC and battery systems, respectively. 
The control vector $u$ is given by 
\begin{align}
         u &  = [u_1, u_2, u_3]^\top := [ v_\mathrm{cm}, I_\mathrm{st}, I_\mathrm{bat}]^\top
\end{align}
where $v_\mathrm{cm}$ is the compressor motor voltage, $I_\mathrm{st}$ is the stack current, and $I_\mathrm{bat}$ is the battery current. 
The FC state $x_\mathrm{fc}$ consists of $p_\mathrm{O_2}$  and $p_\mathrm{N_2}$ representing the oxygen and nitrogen partial pressures in the cathode, respectively, $\omega_\mathrm{cp}$ is the angular speed of the compressor, and $p_\mathrm{sm}$ is the air pressure in the supply manifold. 
The dynamics of the FC system are described by the fist four elements of the vector field $f$ and given by 
\begin{align}
\hspace{-1mm} f_1(x,u) =&  c_1(-x_1 - x_2 + x_4 - c_2) -\frac{c_3 x_1 \psi_\mathrm{ca}(x_1, x_2)}{c_4 x_1 + c_5 x_2 + c_6}\notag \\&-c_7 u_2, \\
\hspace{-1mm} f_2(x) =& c_8(-x_1 - x_2 + x_4 - c_2) - \frac{c_3 x_2 \psi_\mathrm{ca}(x_1, x_2)}{c_4 x_1 + c_5 x_2 + c_6},\\
\hspace{-1mm} f_3(x,u) =& -c_9 x_3 - c_{10} x_3 \left[\left(\frac{x_4}{c_{11}}\right)^{c_{12}} - 1\right] {\psi_\mathrm{cm}}(x_3, x_4) \notag\\&+c_{13} u_1, \\
\hspace{-1mm} f_4(x) =& c_{14} \left[1 + c_{15} \left[\left(\frac{x_4}{c_{11}}\right)^{c_{12}} - 1\right]\right] \notag \\& \times \left[\psi_\mathrm{cm}(x_3, x_4) - c_{16}(-x_1 - x_2 + x_4 - c_2)\right], 
\end{align}
where the meaning and values of the parameters $c_i$ for $i=1,\dots 16$ and the representation of the function $\psi_\mathrm{ca}(x_1,x_2)$ 
can be found in \cite{Talj2009}. 
%
%
The battery state $x_\mathrm{bat}$ consists of $v_\mathrm{soc}$ representing the SOC as a voltage of a fully charged capacitor $C_\mathrm{b}$, and $v_\mathrm{f}$ and $v_\mathrm{s}$ representing voltages in two RC networks in an equivalent circuit model~\cite{Chen2006}. 
%
%
The dynamics of the battery system are described by 
\begin{align}
f_5(x,u) &= -\frac{1}{R_\mathrm{sd} C_\mathrm{b}} x_5 - \frac{1}{C_b} u_3 \\
f_6(x,u) &= -\frac{1}{R_\mathrm{s}(x_5) C_\mathrm{s}(x_5)} x_6 + \frac{1}{C_\mathrm{s}(x_5)} u_3 \\
f_7(x,u) &= -\frac{1}{R_\mathrm{f}(x_5) C_\mathrm{f}(x_5)} x_7 + \frac{1}{C_\mathrm{f}(x_5)} u_3 
\end{align}
where the values of the parameters $R_\mathrm{sd}$ and $C_\mathrm{b}$, and functions $R_\mathrm{s}$, $C_\mathrm{s}$, $R_\mathrm{f}$, and $C_\mathrm{f}$ are found in \cite{Chen2006}. 
Finally, the last state variable $x_8 = q_\mathrm{dis}$ is added to integrate the battery usage:
\begin{align}
   f_8(u) &= u_3. 
\end{align} 



\section{Proposed Energy Management Strategy} \label{sec:method}

This section introduces the proposed hierarchical energy management structure in \secref{sec:hierarchical_structure},  
and presents the design of the online trajectory planner for adaptive battery discharging in \secref{sec:formulation}. 

\subsection{Overall Hierarchical Structure} \label{sec:hierarchical_structure}

\begin{figure}[!t]
    \centering
    \vspace{1mm}
    \includegraphics[width=0.98\linewidth]{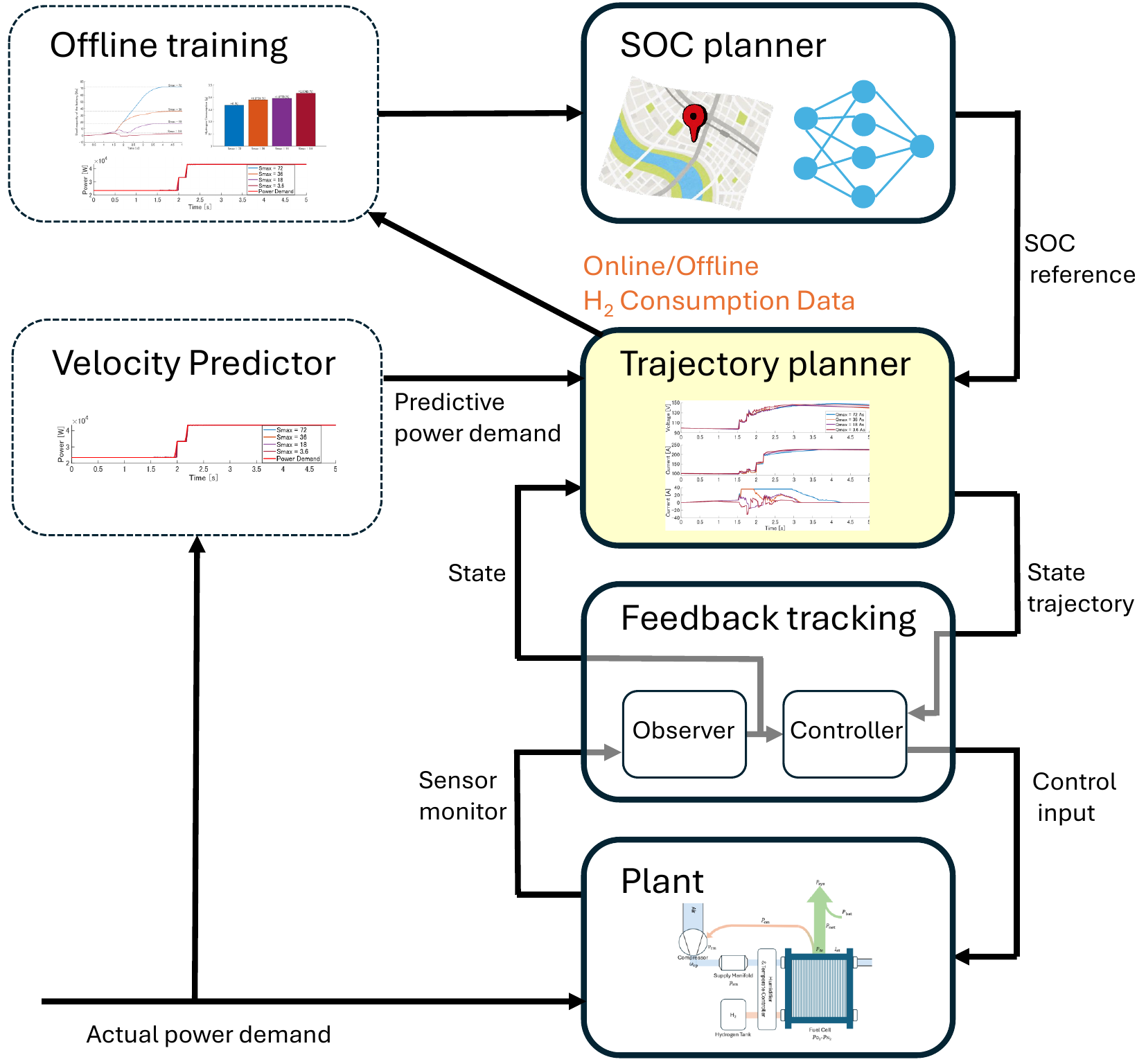}
    \caption{Proposed hierarchical energy management structure}
    \label{fig:control_structure}
\end{figure}

The hierarchical energy management structure proposed in this study is schematically shown in \figref{fig:control_structure}. 
This structure is inspired by recent work on SOC planning using partial trip information \cite{Wei2023,Piras2024} and energy management based on short-term velocity predictor~\cite{Quan2021,Zhou2022}. 
The controller consists of three layers.
At the highest level, an SOC planner determines a battery SOC reference trajectory based on partial trip information. 
In real-world scenarios, obtaining detailed trip information, including precise distance, elevation profiles, and speed profiles, can be challenging or even unattainable. 
However, map service providers may offer simplified information such as road networks, speed limits, and estimated distances between locations~\cite{Piras2024}. 
Neural network based learning approach is considered as a promising candidate for designing the SOC planner that needs to handle intricate relationship between vehicle speed, distance to be covered, and desirable battery depleting schedule \cite{Wei2023,Piras2024}. 
At the second level of the hierarchy, an online trajectory planner generates short-term reference state trajectories considering optimal dynamic coordination between the FC and battery systems. 
It utilizes velocity forecasting for several seconds to minimize hydrogen consumption while tracking the power demand and ensuring safety constraints such as avoiding oxygen starvation and compressor limits. 
To cope with forecasting errors in power demand, online re-planning based on an MPC (receding horizon) strategy is expected to work effectively. 
At the bottom level of the hierarchy, a feedback tracking controller works for achieving the desired state trajectory generated by the planner. 
Also, a state observer is required for estimating the entire state vector $x$, which includes variables that are not directly measurable.  
For this, a nonlinear observer is designed for PEMFC systems in, e.g., ~\cite{Ouyang2017}, and for batteries in, e.g.,~\cite{Ouyang2014}. 

In this paper, we focus on the development of the online trajectory planner at the second layer in \figref{fig:control_structure}. 
The key idea of the design in this paper is that 
the amount of electricity dischargeable from the battery in each driving segment can be explicitly and adaptively specified by the SOC planner.  
With this design, we aim to facilitate the coordination between the SOC planning and trajectory planning in the first two layers in the hierarchy. 
With existing learning methods for the SOC planner~\cite{Wei2023,Piras2024}, training dataset has been generated based on simulations using ECMS as energy management strategy, which does not consider the dynamic responses of the FC system. 
However, considering that the efficiency of the FC system deteriorates during the transients following rapid changes in the power demand due to slow responses of the compressor and the supply manifold~\cite{Vahidi2006}, it would be crucial to take into account the dynamic coordination of the FC and battery systems during the training of the SOC planner. 
To this end, the proposed method enables quantification of the value of discharging a fixed amount of electricity for saving hydrogen consumptions, and these data can be used for training of the SOC planner.  
This is illustrated and further discussed with a numerical example in \secref{sec:simulation} (see \figref{fig:H2_consumption}).

\subsection{Online trajectory planner} \label{sec:formulation}

This paper proposes an online trajectory planning method based on iLQR \cite{Li2004,Tassa2012,Howell2019}, which is one of the state-of-the-art trajectory optimization tools developed for robot motion planning. 
It numerically solves an optimal control problem at each time step and works in an MPC (receding horizon) framework. 
While various solvers are proposed for optimal control problems, direct shooting~\cite{Betts2010} is a popular class of methods, 
and iLQR and its variant, Differential Dynamic Programming, have attracted great attention especially in robotics~\cite{Chen2019,Dantec2022,Wensing2024}, since they are fast and have a low memory footprint, making them amenable to embedded  implementation. 
Although the original iLQR \cite{Li2004,Tassa2012} was not able to handle state and input constraints, the ALTRO algorithm proposed in \cite{Howell2019} combines an augmented Lagrangian method to handle general constraints. 
While some discretization method is required to obtain a discrete-time state equation from the original continuous-time model~\eqref{eq:state_equation}, we used variational-equation based discretization presented in \cite{shigematsu2024}. 

The algorithm of iLQR is iterative. 
It starts from a nominal control sequence $\{ \bar{u}^k \}_{k=0}^{N-1}$ and corresponding nominal trajectory $\{ \bar{x}^k \}_{k=0}^{N}$, where $N$ stands for the outlook horizon. 
Each iteration consists of two processes called the \emph{backward pass}, which solves an optimal control problem for the linearized  dynamics around the nominal trajectory and quadratic approximation of the cost function, i.e., a time-varying LQR problem, based on the dynamic programming principle, and the \emph{forward pass}, which integrates the original nonlinear state equation to update the nominal control sequence $\{ \bar{u}^k \}_{k=0}^{N-1}$ and the nominal trajectory $\{ \bar{x}^k \}_{k=0}^{N}$ while verifying the validity of the optimal solution obtained for the approximated LQR problem. 
The nonlinear optimal control problem used in this paper is given by the following minimization problem where the update of the nominal control sequence $\{ \Delta u^k \}_{k=0}^{N-1}$ with $u^k := \bar{u}^k + \Delta u^k$ is determined at each iteration:
\begin{align}
 &\hspace{-2mm} \min~J = \sum_{k=0}^{N-1}\{l_\mathrm{ref}(x^k,u^k) +l_\mathrm{e}(u^k) +l_\mathrm{s}(\Delta{u^k})  \}\\
 & \mathrm{s.t.} \quad x^0 = x(0), \\
 & \hspace{8mm} x^{k+1} = F(x^k, u^k), \quad \forall k \in \mathcal{N}, \\
 &\hspace{8mm}  \lambda_\mathrm{O_2}(x^k) \ge  \lambda_\mathrm{min},  \quad \forall k \in \mathcal{N}, \label{eq:const_oxygen}\\ 
 &\hspace{8mm} x_2^k \geq a_1\psi_\mathrm{cm}(x_1^k, x_2^k) + b_1,  \quad \forall k \in \mathcal{N} \label{eq:const_choke}, \\
&\hspace{8mm} x_2^k \leq a_2\psi_\mathrm{cm}(x_1^k, x_2^k) + b_2,   \quad \forall k \in \mathcal{N}, \label{eq:const_surge}\\
&\hspace{8mm} v_\mathrm{cm, {min}} \leq u_1^k \leq v_{\mathrm{cm,max}},  \quad \forall k \in \mathcal{N}, \label{eq:const_u1} \\
&\hspace{8mm} I_\mathrm{st, {min}} \leq u_2^k \leq I_{\mathrm{st,max}},  \quad \forall k \in \mathcal{N}, \label{eq:const_u2}  \\
&\hspace{8mm} I_\mathrm{bat,{cmax}} \leq u_3^k  \leq I_\mathrm{bat,{dmax}},  \quad \forall k \in \mathcal{N},  \label{eq:const_u3} \\
&\hspace{8mm} x_8^k \leq Q_\mathrm{max},  \quad \forall k \in \mathcal{N}, \label{eq:const_Qmax}
\end{align}
where $\mathcal{N} := \{0, \dots, N \} $, and $F$ stands for the state transition map derived from the state equation \eqref{eq:state_equation}. 
The objective function consists of three terms of $l_\mathrm{ref}$ for the error between the total output power $P_\mathrm{sys}$ and the demand power $P_\mathrm{ref}$, $l_\mathrm{e}$ for the minimization of the stack current, which is proportional to the hydrogen consumption, and $l_\mathrm{s}$ for improving numerical stability of the iLQR algorithm, and given as 
\begin{align}
    & l_\mathrm{ref}(x^k,u): = W_\mathrm{ref} \{P_\mathrm{sys}^k(x^k,u^k)-P_\mathrm{ref}^k\}^2, \\
    & l_\mathrm{e}(u^k) := W_\mathrm{e} u_2^k,  \\
    & l_\mathrm{s}(\Delta{u^k}) := \Delta{u^k}^\top W_\mathrm{s} \Delta{u^k}, 
\end{align}
where $W_\mathrm{ref}$, $W_\mathrm{e}$, and $W_\mathrm{s}$ are the weights for these terms. 
The constraint \eqref{eq:const_oxygen} is for avoiding oxygen starvation by imposing the lower bound $\lambda_\mathrm{min}$ for the oxygen excess ratio $\lambda_\mathrm{O_2}$. 
The constraints \eqref{eq:const_choke} and \eqref{eq:const_surge} represents the choke and surge boundaries of the compressor, respectively, and the constants $a_1$, $b_1$, $a_2$, and $b_2$ are given in \cite{Vahidi2006}. 
The constrains \eqref{eq:const_u1}--\eqref{eq:const_u3} specifies upper and lower limits of the inputs $u_1$ to $u_3$, respectively. 
The constraint \eqref{eq:const_Qmax} imposes the upper limit on $x_8 = q_\mathrm{dis}$, and this condition is responsible for allowing a fixed amount of electricity $Q_\mathrm{max}$ specified by the upper planner. 
In addition to the above, we added constraints to ensure that $x_1=p_\mathrm{O_2}$ becomes non-negative and that the cathode and supply manifold pressures are larger than the atmospheric pressure.  

\section{Simulation and Discussion} \label{sec:simulation}

This section provides numerical results as a proof of concept example to show the effectiveness of the proposed online trajectory planner.  
The closed-loop behavior of the proposed planner is discussed, and the value of the dynamic coordination of the FC and battery systems is quantified. 
The parameters of the FC and battery systems are based on \cite{Talj2009,Ouyang2017}, and summarized in \tabref{tab:parameters}. 
The sampling period is set to $\Delta t = \SI{0.05}{s}$ and the horizon to $N=10$. 
The weights are $W_\mathrm{ref}=100$, $W_\mathrm{c} = 0.01$, and $W_\mathrm{s}=\mathrm{diag}(1,1,0.01)$, and the lower limit of the oxygen excess ratio is $\lambda_\mathrm{min} = 1.5$. 
With this setting, we compare the results for four cases of discharging amount: $Q_\mathrm{max}=\SI{72}{As}$, $\SI{36}{As}$, $\SI{18}{As}$, and $\SI{3.6}{As}$, which correspond to the amounts of charge such that the battery can be discharged with the maximum current of $\SI{36}{A}$ for $\SI{2}{s}$, $\SI{1}{s}$, $\SI{0.5}{s}$, and $\SI{0.1}{s}$, respectively.

\begin{figure}[!t]
    \centering
    \includegraphics[width=0.96\linewidth]{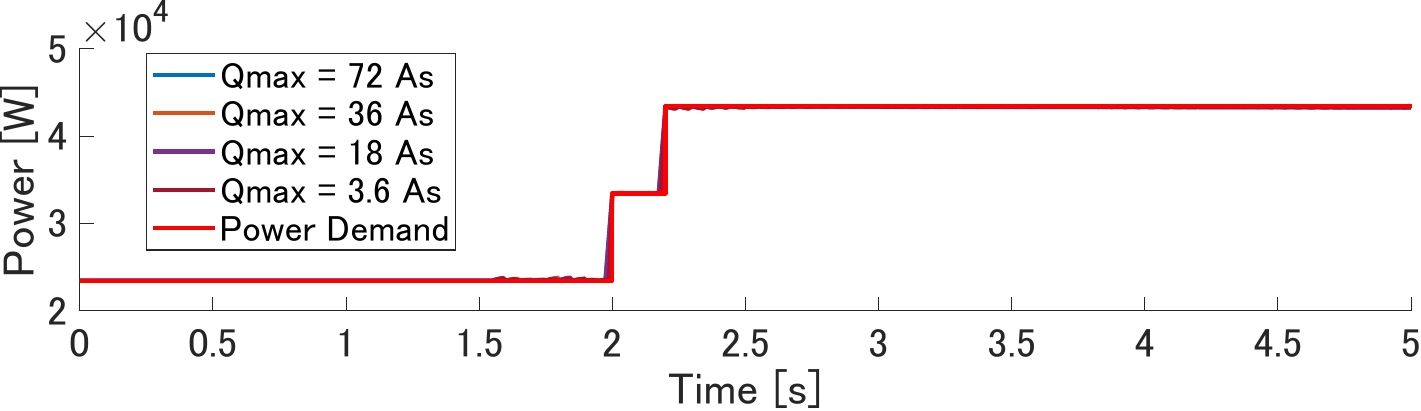}
    \caption{Supplied power $P_\mathrm{sys}$ for each $Q_\mathrm{max}$ and the reference $P_\mathrm{ref}$ }
    \label{fig:demand_power}
\end{figure}

\begin{figure}[!t]
    \centering
    \includegraphics[width=0.98\linewidth]{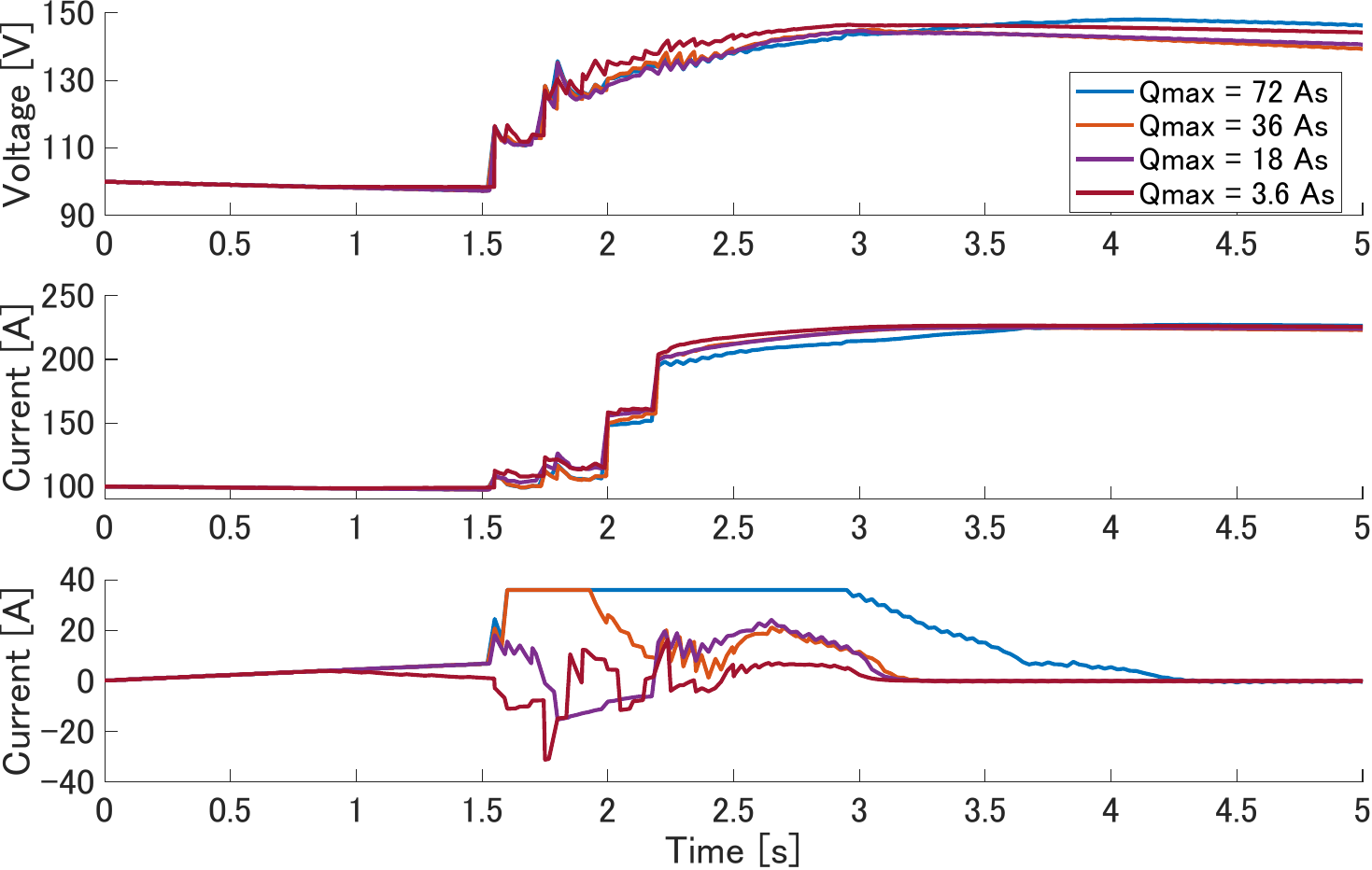}
    \caption{Time courses of the control input $u = [ v_\mathrm{cm},I_\mathrm{st},I_\mathrm{bat}]^\top$}
    \label{fig:input_trajectory}
\end{figure}

An arbitrary test sequence consisting of two steps is considered for the reference of the power demand $P_\mathrm{ref}$, which is shown by the red line in \figref{fig:demand_power}. 
By utilizing the short-term velocity predictor, the controller gains imformation of an upcoming $T_\mathrm{h} = \SI{0.5}{s}$ in advance and can subsequently prepare for the transient responses. 
As shown in \figref{fig:demand_power}, the supplied power $P_\mathrm{sys}$ tracks the power demand  $P_\mathrm{ref}$ without significant error. 
\Figref{fig:input_trajectory} shows the time courses of the planned input variables: the compressor motor voltage $v_\mathrm{cm}$, the stack current $I_\mathrm{st}$, and the battery current $I_\mathrm{bat}$. 
It can be seen that the controller prepares for the step increase in the power demand at $t = \SI{2.0}{s}$ by increasing the compressor voltage $v_\mathrm{cm}$ in advance at around $t = \SI{1.5}{s}$, which increases the oxygen flow to the PEMFC stack. 
As the compressor consumes more power after $t = \SI{1.5}{s}$, the stack current $I_\mathrm{st}$ and/or the battery current $I_\mathrm{bat}$ also increases. 

\begin{table}[!t]
  \centering
  \caption{Parameter settings} \label{tab:parameters}
  \begin{tabular*}{9cm}{@{\extracolsep{\fill}}lll} \hline
Number of cells in fuel cell stack & $n$ & 381 \\
Single stack cathode volume & $V_{\mathrm{ca}}$ & $0.01 \, \mathrm{m}^3$ \\
Supply manifold volume & $V_{\mathrm{sm}}$ & $0.02 \, \mathrm{m}^3$ \\
Mechanical efficiency of motor & $\eta_{\mathrm{cm}}$ & 98\% \\
Compressor efficiency & $\eta_{\mathrm{cp}}$ & 80\% \\
Motor resistance & $R_{\mathrm{cm}}$ & $0.82 \, \Omega$ \\
Compressor and motor inertia & $J_{\mathrm{cp}}$ & $5 \times 10^{-5} \, \mathrm{kg} \cdot \mathrm{m}^2$ \\
Torque sensitivity & $k_t$ & $0.0153 \, \mathrm{N} \cdot \mathrm{m} \cdot \mathrm{A}^{-1}$ \\
Back EMF constant & $k_v$ & $0.0153 \, \mathrm{V} \cdot (\mathrm{rd}/\mathrm{s})^{-1}$ \\
Air specific heat ratio& $\gamma$ & $1.4$ \\
Air specific heat & $C_p$ & $1004 \, \mathrm{J} \cdot (\mathrm{kg} \cdot \mathrm{K})^{-1}$ \\
Cathode output throttle discharge coeff. & $C_D$ & $0.0124$ \\
Cathode output throttle area & $A_T$ & $0.002 \, \mathrm{m}^2$ \\
Cathode inlet orifice & $k_{\mathrm{ca,in}}$ & $\SI{3.629e-6}{kg.s.Pa^{-1}}$ \\
Oxygen molar ratio at cathode inlet & $y_\mathrm{O_2,atm}$ & $0.21$\\
Temperature of the fuel cell stack & $T_{\mathrm{st}}$ & $353.15 \, \mathrm{K}$ \\
Atmospheric temperature & $T_{\mathrm{atm}}$ & $298.15 \, \mathrm{K}$ \\
Average ambient air relative humidity &$\phi_\mathrm{atm}$& $0.5$ \\
Minimum allowed compressor voltage & $v_{\mathrm{cm,min}}$ & $0 \, \mathrm{V}$ \\
Maximum allowed compressor voltage & $v_{\mathrm{cm,max}}$ & $300 \, \mathrm{V}$ \\

Minimum allowed stack current & $I_{\mathrm{st,min}}$ & $0 \, \mathrm{A}$ \\
Maximum allowed stack current & $I_{\mathrm{st,max}}$ & $616 \, \mathrm{A}$ \\
\hline
Number of parallels connections & - & 12 \\
Number of series connections & - & 15 \\
Nominal voltage of the battery & - & $48 \, \mathrm{V}$ \\
Capacity of the battery & - & $12 \, \mathrm{Ah}$ \\
Maximum allowed charging current & $I_{\mathrm{bat,cmax}}$ & $-36 \, \mathrm{A}$ \\
Maximum allowed discharging current & $I_{\mathrm{bat,dmax}}$ & $36 \, \mathrm{A}$ \\
     \hline
  \end{tabular*} \label{tab:input_parameters}
\end{table}

\begin{figure}[!t]
    \centering
    \includegraphics[width=0.9\linewidth]{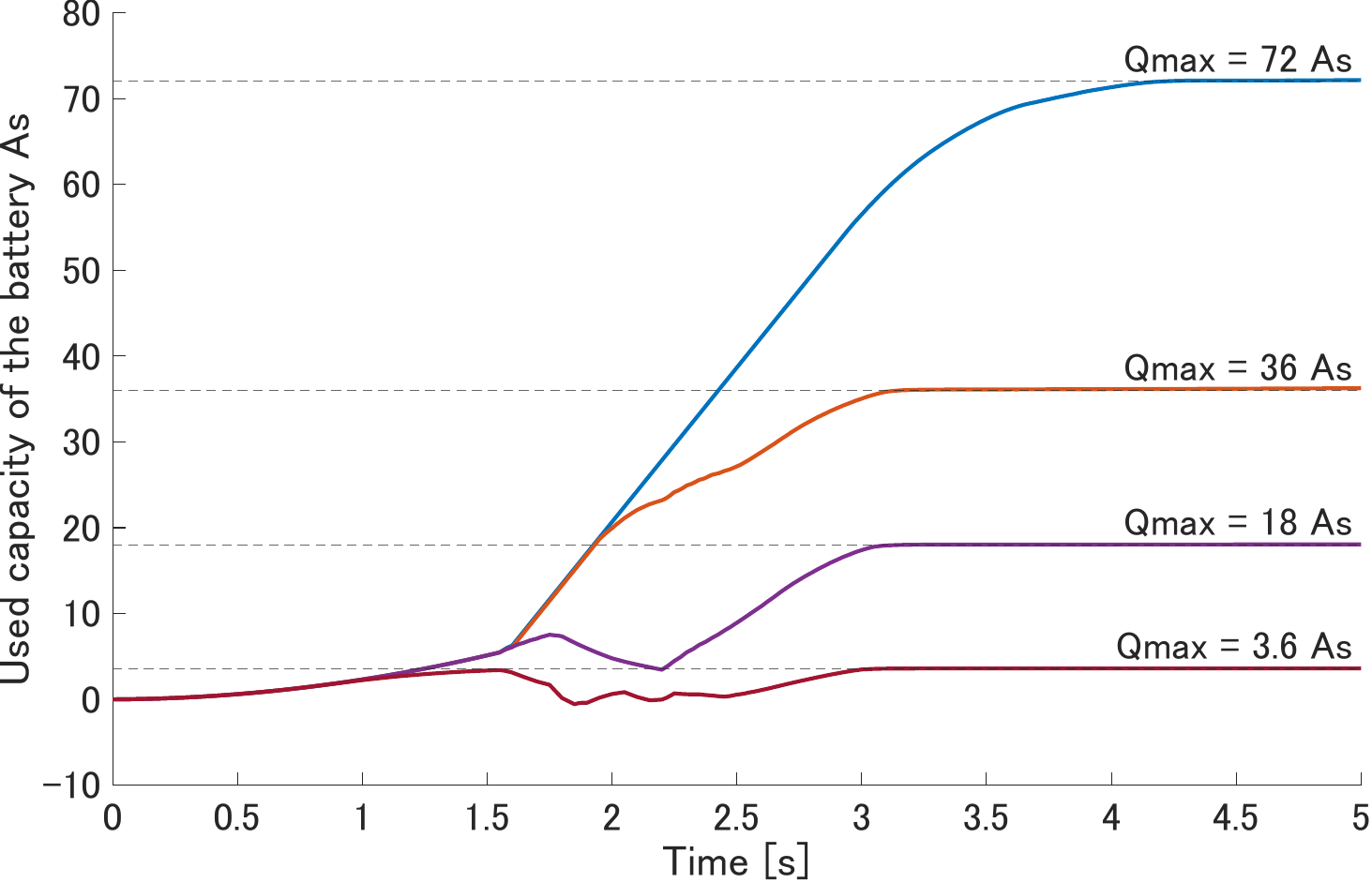}
    \caption{Time course of the state $q_\mathrm{dis}$}
    \label{fig:q_dis}
\end{figure}

To clarify the difference in the battery usage, \figref{fig:q_dis} shows the time courses of the state $q_\mathrm{dis}$ representing the amount of electricity discharged. 
In all cases, $q_\mathrm{dis}$ increases gradually until about $t=\SI{1.5}{s}$. 
After that, with the setting of $Q_\mathrm{max} = \SI{3.6}{As}$, as an overall trend, the battery is charged before the step increase in the demand, and gradually discharged after the increase in the demand. 
The value of $q_\mathrm{dis}$ becomes lowest just before the first increase in the power demand at $t=\SI{2}{s}$, and then rises gradually.
Similarly, with the setting of $Q_\mathrm{max} = \SI{18}{As}$, the value of $q_\mathrm{dis}$ drops near the second increase in the power demand at $t=\SI{2.2}{s}$.  
These behaviors are similar to the battery usage in the charge-sustaining mode in the sense that the battery is charged when the total load is low and discharged when the total load is high. 
In contrast, with $Q_\mathrm{max} = \SI{36}{As}$ and $\SI{72}{As}$, 
the battery is discharged for all the time until the specified amount of electricity is consumed, and the value of $q_\mathrm{dis}$ monotonically increases. 
These behaviors are similar to the operation in the charge-depleting mode. 
The above results show that the proposed planner can generate trajectories corresponding to both charge-depleting and charge-sustaining modes depending on the setting of $Q_\mathrm{max}$. 
In all cases, the value of $q_\mathrm{dis}$ reaches and does not exceed the specified limit of $Q_\mathrm{max}$, meaning that the proposed trajectory planner effectively achieves the dynamic coordination of the FC and battery systems within the specified discharging limit 
in an adaptive manner.


\begin{figure}
    \centering
    \vspace{-1mm}
    \includegraphics[width=0.9\linewidth]{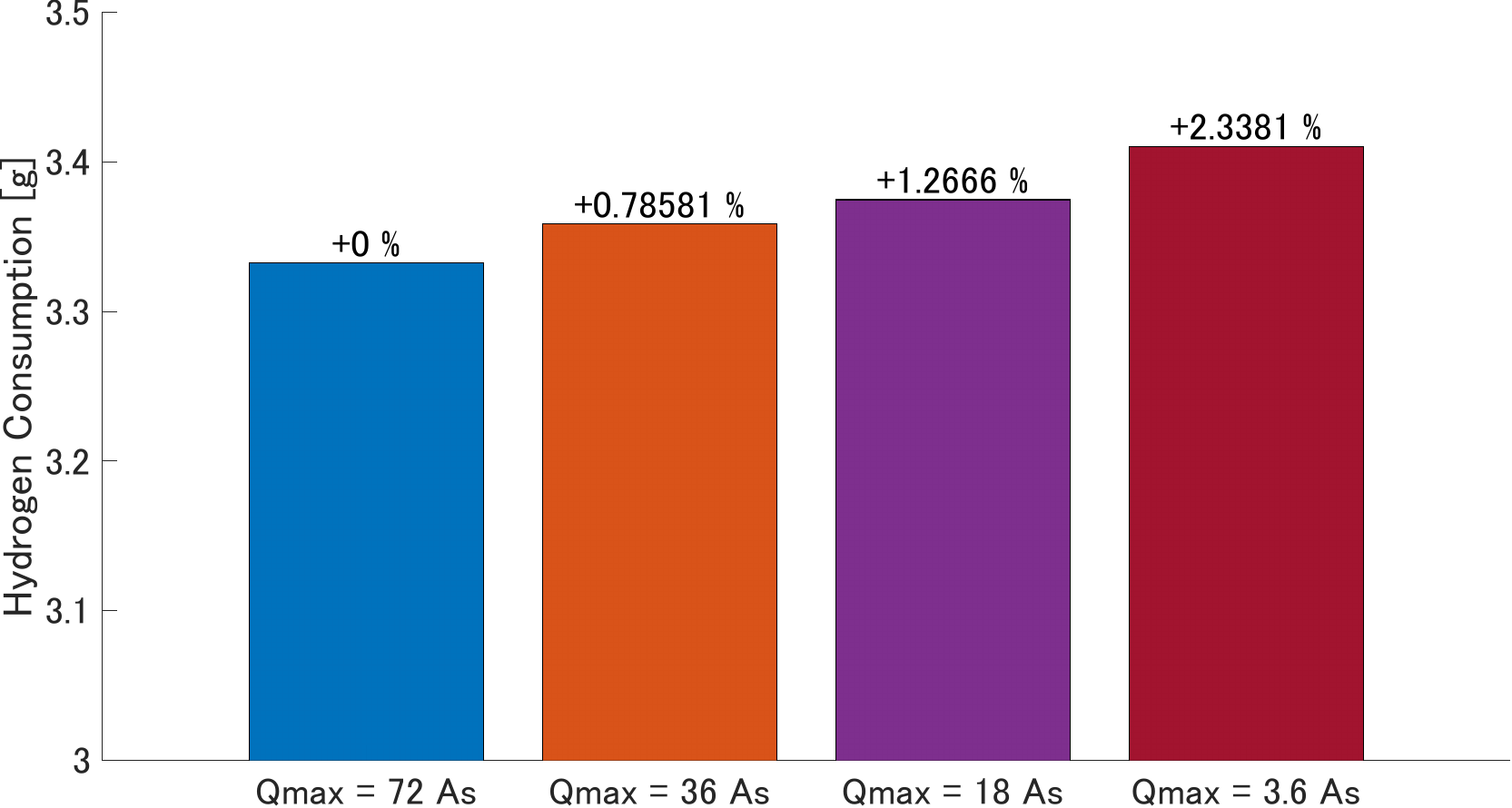}
    \caption{Hydrogen consumption for various setting of $Q_\mathrm{max}$.}
    \label{fig:H2_consumption}
\end{figure}

Finally, \figref{fig:H2_consumption} shows the performance comparison with respect to the hydrogen consumption. 
It can be seen that there is approximately $\SI{2.3}{\%}$ difference between the cases of $Q_\mathrm{max}=\SI{72}{As}$ and $\SI{3.6}{As}$. 
These results quantify the value of discharging a fixed amount of electricity. 
Given the fact that the proposed planner allows to explicitly specify the discharging amount, it is not difficult to gather data on the trade-off between the amount of battery usage and reduction in hydrogen consumption both through offline simulations and online data acquisition that can be used as historical data in the learning process of the SOC planner. 
Thus, the proposed method can be seen as a key component for a better coordination between the first two layers in the hierarchical energy management structure in \figref{fig:control_structure}, where the dynamic performance of the FC and battery systems can be included in the training of the SOC planner.

\section{Conclusions} \label{sec:conclusion}

In this paper, we discussed how to construct a hierarchical energy management structure that is suitable for adaptive SOC trajectory planning under high uncertainties in driving scenarios.
Although existing learning methods for designing an SOC planner effectively incorporate partial trip information, dynamic responses of the FC and battery systems cannot be taken into account. 
For the better coordination between the SOC planner and trajectory optimizer, we proposed an iLQR based online trajectory planning approach where the dischargeable amount of electricity in each driving segment can be explicitly and adaptively specified by the SOC planner.  
Owing to the proposed approach, it becomes easy to gather data on the value of utilizing a fixed amount of battery both through offline simulations and online data acquisition for the training of the SOC planner. 

Future directions of this work include development of a learning method for the SOC planner based on the proposed hierarchical strategy.   
Also, as next steps for real application of the proposed online trajectory planner, implementation of the velocity predictor, the state observer, and low-level controllers are required. A robustness analysis of the proposed planner under prediction errors is crutial to verify the real-life performance. 
It is also necessary to cope with slow variations in model parameters due to temperature and humidity dynamics.

\bibliography{IEAcon2024}
\bibliographystyle{IEEEtran}


\end{document}